\documentclass[aps,prd,twocolumn,superscriptaddress,nofootinbib]{revtex4}

\usepackage{epsfig}
\usepackage{amsmath}
\usepackage{dcolumn}
\usepackage[latin1]{inputenc}

\usepackage{here}

\begin{document}


\title{\textbf{A note on regularization and renormalization}}
\author{M. Stenmark\footnote{email: marten.stenmark@tsl.uu.se}}
\address{Department of Radiation Sciences\\ Uppsala
  University \\ Box 535, S-75121 Uppsala }

\begin{abstract}
We look at sections of a function bundle over the space of linear
differential operators. We find that one can construct an
isomorphism between a certain quotient bundle and the fourier
counterpart of the original bundle defined by formal integration
by parts. We also show that differential renormalization is an
example of this technique.

\end{abstract}
\maketitle
\section{Introduction}
The systematics of renormalization rests on solid ground since
several years. The problem of singular functions that comes from
an ill defined Lagrangian can be taken care of via the procedure
of introducing counter terms. If in addition the Lagrangian is
renomalizable it is enough to add one set of counterterms which
takes care of the infinities to all orders\cite{HOOFT}. In this
paper we will consider certain function bundles whose sections may
be "regularized" with the help of the linear differential operator
in the space of linear operators which it is a section above.
Furthermore the Fourier transform will prove to be a homeomorphism
onto the Fourier counterpart of the bundle. In this bundle the
functions will yield finite results. In the next section we will
show that one may define a quotient bundle that is isomorphic to
the Fourier bundle and study how this comes about. We study why
the results become finite and why this is actually what happens in
differential renormalization.

\section{regularization and renormalization of the function bundle }
Consider the space of linear differential operators, call it
$\xi$. Consider the function algebra consisting of functions with
well defined Fourier transform together with the functions
generated by applying an operator to these. The added functions do
not necessarily have a Fourier transform. This algebra is for each
operator $L\in\xi$ a section over the point $L\in\xi$, call it
$\mathcal{A}_L$. The function bundle we will study is the one
spanned by these sections.

First we define the Fourier transform of a function in
$\mathcal{A}_L$. We do so by formal integration by parts where the
surface terms are neglected. If the function is in the subalgebra
consisting of functions with Fourier transform $\mathcal{B}$ the
surface terms are zero so the only change is for the functions
generated by $L$. For such a function one has $f=Lg=\sum
a_i\partial^i g$ and thus performing formal integration by parts
results in a convolution of the Fourier transform of derivatives
of the operator coefficients $a$ and the fourier transform of $g$.
If the operator consists of only derivatives this gives just
multiplication by a power of p. This map is a homeomorphism
between $\mathcal{A}_L$ and $\mathcal{\widehat{A}}_L$.

If we now construct an ideal $\mathcal{I}_L$ of the algebra we may
take quotient with this ideal and construct a quotient algebra
$\mathcal{A}_L/\mathcal{I}_L$. First we define a character
$\epsilon:\mathcal{B}\rightarrow\mathcal{C}$ by
$\epsilon(f\in\mathcal{B})=\widehat{f}(p)$ where we chose p in the
character such that it coincides with the map $\mathcal{F}$. With
this we may define an ideal of $\mathcal{A}$ as
$\mathcal{I}=\{a(b-\epsilon(b))|b\in\mathcal{B},a\in\mathcal{A}\}$.

It is now strait forward to use the fundamental homeomorphism
theorem \cite{kostrikin} to show that the below diagram commutes.
\begin{eqnarray}
&&\mathcal{A}\rightarrow\mathcal{\widehat{A}}\nonumber
\\
&&\downarrow \hspace{4 mm} \nearrow\nonumber
\\
&&\mathcal{A}/\mathcal{I}
\end{eqnarray}

To prove this we observe that it is enough to show that
$\mathcal{I}$ is in the kernel of the homomorphism between
$\mathcal{A}$ and $\mathcal{\widehat{A}}$. But due to the choice
of p that coincides with $\mathcal{F}$ this is trivial. Thus the
isomorphism between the quotient bundle and the fourier bundle is
induced by the above diagram.

One may argue that the skipping of surface terms is unphysical,
but it turns out that this can be seen as a counterterm procedure
in the following way. Assume that the function is determined from
some Lagrangian and depends only on the square of $x_\mu$. And
that L fulfills the same invariance. The surface terms in n
dimensional euclidian space can be written,
\begin{eqnarray}
&&\int_{S_\epsilon}d\sigma_\mu f(x)\left(\int dx_\mu Lg\right)=\nonumber \\
&&\epsilon^{n-1}\frac{2\pi^{n/2}}{\Gamma(n/2)}f(0)\left(\int
dx_\mu Lg\right)_{x^2=\epsilon^2} +\mathcal{O}(\epsilon)
\end{eqnarray}
where $S_\epsilon$ is the surface of an epsilon ball about the
origin. We assume above that $Lg$ obeys euclidian invariance. This
terms times a constant can be added to the Lagrangian as a
counterterm as in \cite{dzf} and the constant determined to cancel
the surface term in $\mathcal{F}$. Further trouble may still arise
at the boundary of an infinite sphere but in most cases where $f$
is an exponential this will be damped out by oscillations
\cite{dzf}.

The procedure of differential regularization and renormalization
is an example of the above. Take $\phi^4$ theory as an example and
look to the point $(\partial^2,1/x^4)$. In the function bundle one
has a logarithm with a free mass parameter as the function which
$\partial^2$ acts upon to produce $1/x^4$. The free parameter
traces out a path through the section $\mathcal{A}_{\partial^2}$
of the function bundle. The Fourier map takes this path onto a
path through the corresponding Fourier section.

Future work must look into the problem of symmetries, and among
them those which generate Ward identities. Due to the fact that
regularization may invalidate the Ward identities among greens
functions these are only formal for bare Green functions
\cite{Perez}. Although the function bundle contains bare Green
functions it is in no way trivial to show that the map induced by
the diagram will obey these symmetries and thus generate
appropriate Ward identities which hold globally on the bundle.
Symmetries such as gauge symmetries can be proved to be fulfilled
on certain paths through the bundle \cite{dzf,Perez} but to see if
some general results may be obtained further work is necessary.
The cases which has been checked using DR\cite{dzf} obeys
Callan-Symanzik equations for the free parameters which trace out
the path through the sections of the bundle. Generally one will
have different integration constants for each $L\in\xi$. These can
be coupled to each other using modified Ward identities that
couple different paths together. So far this has just been done
for a few examples mentioned earlier.

The bundles presented above gives a new way of viewing
renormalization since we can look at the bundle of renormalizible
functions as a whole. Certainly much work lie ahead if one is to
understand the total structure of this construction, but the
pavement is made and thus one may start the process of
generalizing the ideas which has been set on formal basis during
the past decades.

 \end{document}